\documentclass[10pt,aps,twocolumn,superscriptaddress,preprintnumbers,notitlepage,nofootinbib]{revtex4-1} 
\usepackage{amsmath,amssymb,bbold,graphicx,natbib,bm}
\usepackage{enumerate}
\usepackage{hyperref}
\usepackage{color}
\usepackage{empheq}
\usepackage{slashed}

\newcommand{\beq}{\begin {equation}}  
\newcommand{\eeq}{\end   {equation}} 
\newcommand{\bea}{\begin {eqnarray}} 
\newcommand{\eea}{\end   {eqnarray}}  
\newcommand{\baa}{\begin {array}   } 
\newcommand{\eaa}{\end   {array}   }     
\newcommand{\bit}{\begin {itemize} }
\newcommand{\eit}{\end   {itemize} }
\newcommand{\be }{\begin {equation}} 
\newcommand{\ee }{\end   {equation}}
\newcommand{\nn }{\nonumber        }

\begin{document}


\preprint{UTTG-24-14}
\preprint{TCC-030-14}

\title{Orbifold Grand Unification: A Solution to the Doublet-Triplet Problem}

\author{Bei Jia}
\author{Jiang-Hao Yu}
\affiliation{Theory Group, Department of Physics and Texas Cosmology Center,
\\The University of Texas at Austin,  Austin, TX 78712 U.S.A.}


\begin{abstract}
To solve the doublet-triplet splitting problem in $SU(5)$ grand unified theories, we propose a four dimensional orbifold grand unified theory by acting $\mathbf{Z}_2$ on the $SU(5)$ gauge group. 
Without an adjoint Higgs, the orbifold procedure breaks the $SU(5)$ gauge symmetry down to the standard model gauge group, and projects the triplet component of the fundamental $SU(5)$ Higgs out. 
In the supersymmetric framework, we show that the orbifold procedure removes two triplet superfields of the Higgs multiplets and leaves us with the minimal supersymmetric standard model, which also solves the hierarchy problem and realizes gauge coupling unification. 
We also discuss possible UV completions of the orbifold theories.

\end{abstract}

\maketitle


\section{Introduction}
\label{sec:intro}

For a long time, Grand Unified Theories (GUTs) have attracted lots of attention because of their theoretical simplicity and beauty~\cite{Georgi:1974sy, Agashe:2014kda}. 
GUTs lead to a beautiful unification among the three forces and among the quarks and leptons in the Standard Model (SM). 
There are more advantages in supersymmetric (SUSY) versions of GUTs.  In SUSY GUT, the so-called hierarchy problem is solved, and gauge coupling unification is truly realized.
Furthermore, the GUT scale, which is the scale of gauge coupling unification, is not far from the Planck scale. 
This helps to suppress the proton decay.

However, there are some known problems in GUTs. The most serious one is the so-called doublet-triplet splitting problem~\cite{Agashe:2014kda}. 
The doublet-triplet splitting problem is the question of why the Higgs doublets are of the order of the weak scale while the colored triplets have to be of the order of the GUT scale in order to obtain unification of couplings and to avoid proton decay.
In SUSY GUT, on top of the usual dimension-6 heavy gauge boson and gaugino mediated proton decay, there are potentially much more dangerous dimension-5 proton decay coming from the exchange of the heavy color triplet Higgs. 
There are many possible ways to address the doublet-triplet splitting problem. 
Several mechanisms for natural doublet-triplet splitting have been suggested, such as the sliding singlet~\cite{Witten:1981kv}, missing partner or missing VEV~\cite{Altarelli:2000fu, Masiero:1982fe, Grinstein:1982um, Babu:1994dq, Babu:1994kb}, and pseudo-Nambu-Goldstone boson mechanisms~\cite{Barbieri:1992yy, Berezhiani:1995sb}.
In models with extra dimension~\cite{Kawamura:1999nj, Kawamura:2000ev, Altarelli:2001qj, Hall:2001pg, Hebecker:2001wq, Hebecker:2001jb}, geometric orbifolds were used to break the bulk GUT symmetry and projects zero modes of the Higgs triplet out of the theory through suitable boundary conditions. Therefore, the doublet-triplet splitting problem is avoided.

In this paper, we try to solve this problem by adopting the orbifold procedure in purely four dimensional GUTs, inspired by the recent developments of orbifold Higgs models~\cite{Craig:2014aea, Craig:2014roa}.
We focus on the $SU(5)$ GUTs, and obtain an orbiforded theory by choosing suitable actions of $\mathbf{Z}_2$ on the gauge group $SU(5)$.
Our four dimensional orbifold GUT is very simple, compared to the five dimensional versions. 
In five dimensional orbifold GUTs, one has to choose proper boundary conditions on branes to remove zero modes of some bulk fields.
While in our four dimensional models, the GUT symmetry is broken by projecting some states out.  
We also comment on the possibility that five dimensional bulk-brane setup could be viewed as one of the UV completions of our four dimensional orbifold GUTs.

This paper is organized as follows. In sec.~\ref{sec:general}, we briefly describe the procedure of field theoretic orbifold in four dimensional gauge theories. In sec.~\ref{sec:orbifold}, the four dimensional $SU(5)$ orbifold GUT is investigated in detail.
In the following section, this orbifold GUT is extended to the minimal supersymmetric case. 
Then we discuss possible UV completions in sec.~\ref{sec:UV}. Finally we conclude our paper in sec~\ref{sec:conclusion}.


\section{Field Theoretic Orbifold}
\label{sec:general}

The word {\it orbifold} comes from string theory, which requires ten spacetime dimensions. One of the simplest way to hide these extra dimensions is to construct an orbifold, which means one constructs a quotient space of the form $\mathbf{R}^6 / \Gamma$, where $\Gamma$ is a subgroup of $SO(6)$. At low energies,
one obtains a four dimensional effective supergravity from the ten dimensional string theory. Along these lines, one can include D3-branes in this setup, and
the effect of orbifolding results in a four dimensional quantum field theory from the worldvolume theory of those D3-branes \cite{KS}. Different choice of the orbifold group will produce different ``daughter'' theories on D-brane worldvolume. This procedure on D-branes is usually called ``field theory orbifold'' \cite{field-orb,Schmaltz:1998bg}, which is what we use in this paper.

From quantum field theory point of view, field theory orbifolding means one starts with a ``mother'' theory, and chooses a discrete group together with its actions on the gauge and global symmetry groups of the mother theory, then keeps those field components that are invariant under these actions. The ``daughter'' theory one obtains has a Lagrangian that is derived from the Lagrangian of the mother theory, via keeping the terms that involves the invariant components only.

The explicit expression of field theoretic obifold procedure goes as follows. We start with a four dimensional mother theory with gauge group $G$, together with an action of a finite group $\Gamma$ (called orbifold group) on $G$.\footnote{If there are global symmetries, then in principle orbifold group can act on those global symmetry groups as well. However, we will not encounter this situation here.} Then various representations of $G$ can be decomposed into direct sums of irreducible representations of $\Gamma$. For us, $G = SU(n)$,  and we will use a decomposition of the fundamental representation of $SU(n)$:
\bea
\mathbf{C}^n = \oplus_i \left( \mathbf{C}^{n_i} \otimes \mathbf{r}_i \right)
\eea
where $\mathbf{C}^n$ denotes $n$ dimensional complex vector space and $\mathbf{r}_i$ are irreducible representations of $\Gamma$. Effectively, we obtain a homomorphism $f: \ \Gamma \rightarrow SU(n)$. Then the action of $\Gamma$ on $SU(n)$ can be viewed as matrix multiplications
\bea
F \mapsto f(\gamma)F, \ \gamma \in \Gamma
\eea
for any field $F$ in the fundamental representation of $SU(n)$. Then field theoretic obifolding means to keep only those field components that are invariant under this action.

This induces a decomposition of the adjoint representation of $SU(n)$, as $\mathbf{C}^n \otimes \mathbf{C}^{n*} = \text{\textbf{adj}} \oplus \mathbf{1}$. Concretely, we have 
\bea
A \mapsto f(\gamma) A f(\gamma)^{\dagger}, \  \gamma \in \Gamma
\eea
for any field $A$ in the adjoint representation of $SU(n)$. Again, we keep only those field components that are invariant under this action.


\section{Orbifold $SU(5)$ GUT}
\label{sec:orbifold}

In our setup, we start from a mother theory: four dimensional $SU(5)$ GUT without an adjoint Higgs. 
The matter content with the familiar decomposition of $SU(5)$ representations under the $SU(3) \times SU(2)$ symmetry is:
\bea
	A_\mu^a : \ \mathbf{24}     &=& (\mathbf{8}, \mathbf{1}) \oplus (\mathbf{1}, \mathbf{3}) \oplus (\mathbf{1},\mathbf{1}) \oplus (\mathbf{3},\mathbf{2}) \oplus (\bar{\mathbf{3}}, \mathbf{2}), \nn\\ 
	H^*,  \psi : \ \bar{\mathbf{5}} &=& (\bar{\mathbf{3}}, \mathbf{1}) \oplus (\mathbf{1},\mathbf{2}), \\
	\Psi : \ \mathbf{10}      &=& (\mathbf{3}, \mathbf{2}) \oplus (\bar{\mathbf{3}}, \mathbf{1}) \oplus (\mathbf{1},\mathbf{1}). \nn
\eea
%
The Lagrangian can be  written as 
\bea
{\mathcal L} &=& 	-\frac14 F^a_{\mu\nu}F^{a\mu\nu} + (D^\mu H)^\dagger D_\mu H - V(H^2) \nn\\
   && + \bar{\psi}  \gamma^\mu D_\mu  \psi + 
	{\rm Tr}(\bar{\Psi}   \gamma^\mu D_\mu  \Psi)\nn\\
	&&+
	y_5 \bar{\psi} \Psi  H^* 
	+ y_{10} \bar{\Psi} \Psi  H + h.c.,
\eea
where the Higgs potential is $V(H^2) = - \mu^2 H^2 + \lambda H^4$, and 
the covariant derivatives are
\bea
	D_\mu  \psi &=& \left(\partial_\mu + i g_U A_\mu\right)\psi , \nn\\
	D_\mu  \Psi &=& \partial_\mu \Psi + i g_U A_\mu\Psi + i g_U \Psi A_\mu^T,
\eea
where $g_U$ is the $SU(5)$ gauge coupling.
Note that there is no adjoint Higgs field to break $SU(5)$ symmetry. 
We will use the orbifold procedure to break $SU(5)$ down to  $SU(3) \times SU(2) \times U(1)$ 
symmetry~\footnote{This $U(1)$ may not be directly identified as the SM hypercharge. However, as discussed in \cite{Craig:2014aea, Craig:2014roa}, one could incorporate extra $U(1)$s to obtain a linear combination, which could work as the SM hypercharge.}.

For our purposes, we will be focusing on a simple $\mathbf{Z}_2$ orbifold of the mother GUT theory. To perform the orbifold procedure, we first have to specify actions of $\mathbf{Z}_2$ on the $SU(5)$ gauge group. This can be achieved decomposing the fundamental representation of $SU(5)$ as 
\bea
\mathbf{C}^5 = (\mathbf{C}^2 \otimes \mathbf{r}_1) \oplus (\mathbf{C}^3 \oplus \mathbf{r}_2),
\eea
where $\mathbf{r}_1$ is the trivial representation of $\mathbf{Z}_2$, while $\mathbf{r}_2$ is the only nontrivial irreducible representation of $\mathbf{Z}_2$. Effectively, this can be viewed as mapping the elements $\gamma_1, \gamma_2$ of $\mathbf{Z}_2$ into the fundamental representation of $SU(5)$:
\bea
f: \ \gamma_1 \mapsto \left( \mathbf{1}_{5} \right), \gamma_2 \mapsto \left( \begin{array}{ccc|cc}
&&& \\ & -\mathbf{1}_{3} & & 0 \\
&&& \\ \hline & 0 & & \mathbf{1}_{2} \end{array}
\right),
\eea
where $\mathbf{1}_{n}$ is the $n\times n$ unit matrix.
The fundamental Higgs $H$ and adjoint gauge fields $A$ transform as the following:
\bea
	 H \to f(\gamma_{i}) H , \ \ A \to  f(\gamma_{i}) A f(\gamma_{i})^\dagger, \ i = 1, 2.
\eea
We demand that there is no $\mathbf{Z}_2$ transformation on the fermion fields\footnote{This assumption can be justified by possible UV completions, such as geometric orbifolds from extra dimensions.}. 
The orbifold procedure then extracts the $\mathbf{Z}_2$ invariant components of these fields: 
\begin{itemize}
\item The triplet component of the Higgs is not invariant, therefore removed from our theory.
Only the doublet component of the Higgs is left after orbifold, and we identify it as the SM Higgs doublet $H_2$.
\item Similarly, the $X$ and $Y$ components of the $SU(5)$ gauge bosons are removed. We are left with gauge fields of $SU(3) \times SU(2) \times U(1)$.
\end{itemize}
So after orbifolding, we have only Standard Model fields left, and obtain the Lagrangian of the daughter theory:
the standard model with all the gauge couplings the same at the GUT scale. The Yukawa terms are explicitly listed as follow:
\bea
	{\mathcal L}_{Yuk} =  y_5 (\bar{\ell} H_2 e_R + \bar{q} H_2 d_R )
	+ y_{10} \bar{q} \tilde{H}_2 u_R + h.c..
\eea

If we take the orbifold procedure as from certain possible geometric orbifold of extra dimensions, and take the orbifold scale as the GUT scale,
all the removed fields are above the GUT scale. 
Thus proton decay problem and the doublet-triplet splitting problem are solved. 
On the other hand, although all the gauge couplings are unified at the GUT scale, it predicts the wrong gauge couplings at the $M_Z$ scale. In another word, those three gauge couplings of the SM do not unify together if one evolutes from electroweak scale to the GUT scale. 
Furthermore, due to proton decay constraints the resulting neutrino masses turn out to be too small.
In the following section, we will extend our discussion to the SUSY version of our orbifold GUT. This could solve all the problems that are mentioned above.


\section{Orbifold SUSY $SU(5)$ GUT}
\label{sec:SUSY}

To realize gauge coupling unification, it is better to study the SUSY extension of our obifold $SU(5)$ GUT.
In SUSY framework, there are additional benefits. Due to radiative symmetry breaking in the Higgs potential, the hierarchy problem is naturally solved in SUSY GUT. 
However, the doublet-triplet splitting problem still exists. 
We will perform a similar orbifold procedure in a SUSY GUT theory to solve this doublet-splitting problem.

We will focus on the minimal supersymmetric $SU(5)$ GUT, but without the GUT Higgs filed\footnote{Because the adjoint representation of $SU(n)$ does not contribute to anomaly coefficient, our model is free of gauge anomaly.}, because we will break $SU(5)$ via orbifolding as in the last section. The superfield content of our mother SUSY $SU(5)$ GUT is 
\begin{center}
\begin{tabular}{ c | c | c | c | c }
  $V$ & $H_u$ & $H_d$ & $\Phi_{\bar{\mathbf{5}}}$ & $\Phi_{\mathbf{10}}$ \\ \hline
  \textbf{adj} & $\mathbf{5}$ & $\bar{\mathbf{5}}$ & $\bar{\mathbf{5}}$ & $\mathbf{10}$ \\
\end{tabular}
\end{center}
with the following superspace Lagrangian
\bea
\mathcal{L} &=& \int d^4\theta \left( \Phi_{\bar{\mathbf{5}}}^{\dagger} e^{2g_{U}V^a} \Phi_{\bar{\mathbf{5}}} + \Phi_{\mathbf{10}} e^{2g_{U}V^a} \Phi_{\mathbf{10}} \right) \nn \\
&& \int d^4\theta \left( H_u^{\dagger} e^{2g_{U}V^a} H_u + H_d^{\dagger} e^{2g_{U}V^a} H_d \right) \nn \\
&& + \int d^2\theta \left( W^a W^a + \mu H_u H_d\right) + h.c. \nn \\
&& + \int d^2\theta \left( f_u H_u \Phi_{\mathbf{10}} \Phi_{\mathbf{10}} + f_d H_d \Phi_{\bar{\mathbf{5}}} \Phi_{\mathbf{10}}\right) + h.c. \nn \\
&& + \ \text{soft terms}.
\eea
Similar to the non-SUSY version, we don't need the adjoint Higgs multiplet to break the $SU(5)$ symmetry.

Let's now apply the simple $\mathbf{Z}_2$ orbifolding from the last section, to the vector superfield $V$ as well as the two Higgs chiral superfileds $H_u$ and $H_d$. We still demand that there is no $\mathbf{Z}_2$ action on the other fields in our mother theory. Schematically, we have 
\bea
	 H_{u,d} \to f(\gamma_{i}) H_{u,d} , \ \ V \to  f(\gamma_{i}) V f(\gamma_{i})^\dagger, \ i = 1, 2.
\eea
Note that because we have included soft terms in our theory, SUSY is explicitly broken, and there is no R-symmetry. Therefore, there is no continuous global symmetry on which $\mathbf{Z}_2$ can act. Hence after orbifolding, the invariant components of the $SU(5)$ vector multiplet form vector multiplets of $SU(3) \times SU(2) \times U(1)$. In addition, only the doublet components of the Higgs chiral superfileds remains, thus solving the doublet-triplet splitting problem.

Finally by keeping the terms that contains only those invariant in the above Lagrangian, we end up with the superspace Lagrangian of the minimal supersymmetric standard model (MSSM) at the compactification scale.
Therefore, the daughter theory is simply the MSSM, which has all the advantages we discussed above. 


\section{UV Completions}
\label{sec:UV}

There could be many possible UV completions of our simple four dimensional models. One possible solution is to include more than four spacetime dimensions, i.e. extra dimensional models. Then the effective orbifolding procedure we used in this paper can be viewed as a true geometric orbifolding on the extra dimensions. We will discuss some simple five dimensional models along these lines in a moment. Generally speaking, one can hope to embed this setup to something like superstring theory, where orbifolds play an important role. Furthermore, there could be more complicated effects, from string theory, that effectively produce the four dimensional orbifolding procedure we have been using.


A straightforward way to obtain the UV completion is through the geometric interpretation of the orbifold in a GUT model with 
extra dimensions. 
%
A direct UV completion of our four dimensional orbifold $SU(5)$ GUT is to have the $SU(5)$ gauge fields and fundamental Higgs are in the bulk,  while keeping the $\bar{5}$ and $10$ fermions on a four dimensional brane~\cite{Kawamura:1999nj, Kawamura:2000ev, Altarelli:2001qj, Hall:2001pg, Hebecker:2001wq}.
%
Through the orbifold procedure on the boundaries, only some components of the bulk fields have zero modes.   
Because the fermions live on the brane, all the components of the fermion fields are kept.
This explains in four dimensional orbifold GUT only the gauge boson and Higgs field are orbifolded but not the fermion fields.

In the non-SUSY five dimensional GUT, the spacetime is assumed to be factorized into a product of four dimensional Minkowski spacetime $M^4$ and the orbifold $S_1/\mathbf{Z}_2$. The orbifold $S_1/\mathbf{Z}_2$ is obtained by dividing a circle $S_1$ with radius $R$ with a $\mathbf{Z}_2$ transformation which acts on $S_1$ by $x_5 \to - x_5$, where $x_5$ denote the 5-th dimension coordinate. In the bulk, gauge fields have the following  $\mathbf{Z}_2$ parities at one of the boundaries
\begin{equation}
A_\mu^a \to \left( \begin{array}{ccc|c}
&&& \\ &  +  & & - \\
&&& \\ \hline & - & & + \end{array}
\right), \ \ A_5^a \to \left( \begin{array}{ccc|c}
&&& \\ & - & & + \\
&&& \\ \hline & + & & - \end{array}
\right).
\label{SU5BC}
\end{equation}
In order to solve the doublet triplet
splitting problem,
one needs to require that the Higgs has $\mathbf{Z}_2$
parity at one of the boundaries:
\begin{equation}
H \to \left( \begin{array}{c} \\ - \\ \\ \hline + \end{array} \right).
\end{equation}
Applying these boundary conditions, we obtain that only $SU(3) \times SU(2) \times U(1)$ gauge fields
and the doublet Higgs have zero mode. 
The triplet Higgs only appear as the Kaluza-Klein excitation modes, which naturally causes doublet-triplet splitting.

To solve the hierarchy problem, an anti-deSitter (AdS) space metric could be applied:
\bea
	d s^2 = e^{-2 |k|x_5} \eta_{\mu\nu} d x^\mu d x^\nu + d x_5^2,
\eea
where $\eta_{\mu\nu} = {\rm diag}(-1,,1,1,1)$ and $k$ is inverse of the radius of AdS curvature $R$. 
The $S_1/\mathbf{Z}_2$ orbifold fixed points $x_5 = 0$  and $x_5 = \pi R$ are identified as the Planck brane and TeV brane, respectively. 
Due to the AdS warping, an exponential hierarchy between the Planck and TeV scales is generated. The Higgs is localized around the TeV brane, thus the hierarchy problem is solved  in AdS setup.
This scenario has been discussed in Ref.~\cite{Randall:2001gc,Agashe:2002pr}.

A similar procedure along these lines was applied to five dimensional $\mathcal{N} = 1$ SUSY $SU(5)$ GUT in Ref.~\cite{Kawamura:2000ev,Altarelli:2001qj,Hall:2001pg,Hebecker:2001wq}.
In these models, both the GUT group and five dimensional SUSY (corresponding to $\mathcal{N} = 2$ SUSY in four dimension) are broken down to a $\mathcal{N} =1$ supersymmetric models with SM gauge group by compactification on $S^1 / (\mathbf{Z}_2 \times \mathbf{Z}'_2)$.
The bulk particles with $(+, +)$ boundary condition under $\mathbf{Z}_2 \times \mathbf{Z}'_2$ have  zero-modes, while particles with $(+, -)$ or $(-, +)$ boundary conditions have no zero-modes. 
By choosing appropriate boundary conditions on $SU(5)$ supermultiplets, the zero modes contain the necessary field content of the four dimensional $\mathcal{N} = 1$ SUSY $SU(5)$ GUT. 
This scenario has also been extended to warped SUSY $SU(5)$ GUT~\cite{Pomarol:2000hp,Goldberger:2002pc}.
Beside the SUSY $SU(5)$, the orbifold breaking procedure in five dimensional theories was discussed in general gauge groups~\cite{Hebecker:2001jb}, and also specifically $SO(10), E_6$ groups~\cite{Dermisek:2001hp, Kim:2002im}.  
%
%
There could be a corresponding four dimensional effective orbifold procedure on these five dimensional SUSY GUT theories.


\section{Conclusion}
\label{sec:conclusion}

We have studied $SU(5)$ GUT mother theories which do not include any adjoint Higgs, together with actions of a discrete $\mathbf{Z}_2$ on $SU(5)$ to obtain daughter theories, both in non-supersymmetric and supersymmetric cases, via field theoretic orbifolds.
In the non-supersymmetric case, this effective orbifolding projects the triplet Higgs and the gauge bosons $X$ and $Y$ out of the theory, hence solving the doublet-triplet splitting problem as well as avoiding the proton decay.
The daughter theories only contain SM fields with the same gauge coupling strength at the GUT scale. 
%
%
To realize gauge coupling unification at the GUT scale and solve the hierarchy problem, we extend our $SU(5)$ orbifold GUT into supersymmetric framework. 
%
Similar to non-SUSY model, the orbifold procedure is applied to project the $X$ and $Y$ components of gauge boson multiplets the triplet components of Higgs supermultiplets out. 
We obtain the usual MSSM Lagrangian with gauge coupling unification as the daughter theory. 
%
%
Our four dimensional orbifold setup could have possible UV completions. 
We discussed the geometric setup of our orbifold GUT via five dimensional geometric orbifolds, leaving more general discussion to future work.

This orbifold procedure could be applied to other GUT models, such as $SO(10)$, $E_6$ etc. In order to do this, one would have to choose an appropriate orbifold group $\Gamma$, with suitable actions on the gauge group of the mother theory (and possibly on other global symmetry groups if there is any). The details are more complicated, as representations of larger gauge groups are more complicated. We leave this to future work.


\section*{Acknowledgements}

We would like to thank Nathaniel Craig and Can Kilic for helpful discussion. 
The research of the authors is supported by the National Science Foundation under Grant Numbers PHY-1315983 and PHY-1316033.


\end{document}